\documentclass[aps,prd,twocolumn,groupedaddress,amsfonts,amssymb,amsmath,showpacs,showkeys]{revtex4-1}
\usepackage{graphicx}
\usepackage{graphicx}  
\usepackage{dcolumn}   
\usepackage{bm}        
\usepackage{amssymb}   

\begin{document}

\title{Measuring social complexity and the emergence of cooperation from entropic principles}

\author{O L\'opez-Corona$^{1,2,3*}$, P Padilla $^{4}$, A Huerta $^{5}$, D Mustri-Trejo$^{6}$, K Perez$^{2}$, A Ruiz$^{2}$, O Vald\'es$^{2}$ and F Zamudio $^{7}$}
\affiliation{$^{1}$Instituto de Investigaci\'on sobre Desarrollo Sustentable y Equidad Social, Universidad Iberoamericana, Prolongaci\'on Paseo de la Reforma 880, Lomas de Santa Fe, CDMX, M\'exico}
\affiliation{$^{2}$Centro de Ciencias de la Complejidad, Universidad Nacional Aut\'onoma de M\'exico Circuito Escolar, Cd. Universitaria, CDMX, M\'exico}
\affiliation{$^{3}$Former: C\'atedras CONACyT, Universidad Aut\'onoma Chapingo, Carretera M\'exico-Texcoco Kil\'ometro 38.5, Estado de M\'exico, M\'exico}

\affiliation{$^{4}$Instituo de Investigaci\'on en Matem\'aticas Aplicades y en Sistemas, Universidad Nacional Aut\'onoma de M\'exico Circuito Escolar, Cd. Universitaria, CDMX, M\'exico}

\affiliation{$^{5}$Facultad de Ciencias, Universidad Nacional Aut\'onoma de M\'exico Circuito Escolar, Cd. Universitaria, CDMX,  M\'exico}

\affiliation{$^{6}$Posgrado en Ciencias F\'isicas, Universidad Veracruzana, Xalapa, M\'exico}

\affiliation{$^{7}$Grupo de Estad\'istica Social, Universidad Aut\'onoma Chapingo, Carretera M\'exico-Texcoco Kil\'ometro 38.5, Estado de M\'exico, M\'exico}

\date{\today}

\date{\today}

\begin{abstract}
  Assessing quantitatively the state and dynamics of a social system is a very difficult problem. It is of great importance for both practical and theoretical reasons such as establishing the efficiency of social action programs, detecting possible community needs or allocating resources. In this paper we propose a new general theoretical framework for the study of social complexity, based on the relation of complexity and entropy in combination with evolutionary dynamics to asses the dynamics of the system. Imposing the second law of thermodynamics, we study the conditions under which cooperation emerges and demonstrate that it depends of relative importance of local and global fitness. As cooperation is a central concept in sustainability, this thermodynamic-informational approach allows new insights and means to asses it using the concept of Helmholtz free energy. Finally we introduce a new set of equations that consider the more general case where the social system change both in time and space, and relate our findings to sustainability.
\end{abstract}


\maketitle


In general, Complexity comes from the Latin plexus, which means interwoven, thus something complex is difficult to separate because interactions determinate in part the future of the system \cite{Gershenson-2013}.

Because interactions generate novel information, a reductionist scientific approach 
has been recognized to be not appropriate for studying complex systems, due to its
attempts to simplify and separate each component in order to predict its behavior \cite{Gershenson-2013}.
Considering that the information generated by the interactions are not included in the initial and
boundary conditions, the predictability of the system is restricted by Wolfram's computational irreducibility \cite{Wolfram-2002}. 

Interactions can also be used by components to self-organize, producing global patterns from
local dynamics. Furthermore, another major source of complexity is the fact that interactions themselves may change in time, generating non-stationary state spaces. Therefore, even when a solution may be calculated, it will turn inapplicable, for the problem from which it was obtained, does not longer exists \cite{Gershenson-2013}.

Just as Lyapunov exponents characterize different dynamical regimes, or temperature
represents a statistical average of the kinetic energy regarding the nature of the system, it would
be very useful to have global measures for complexity. The interactions between the
components in each particular case are difficult to calculate, but complexity measures that
represent the type of interactions between them, could actually be calculated.

An useful measure of complexity should enable us to answer questions such as: Is a desert more
or less complex than a tundra? What is the complexity of different influenza outbreaks? Which
organisms are more complex: predators or preys; parasites or hosts; individual or social? What
is the complexity of different music genres? \cite{Fernandez-2014}

As Michaelian (2000) \cite{Michaelian-2000} has pointed out, one of the greatest misconceptions preventing the description of biological or social systems from a physical perspective, had its origin in the concept of entropy, and the
formulation of the second law of Thermodynamics during the original development of the theory
of heat engines. In this context, entropy is a measure of the energy that is not available to
perform work. Boltzmann later gave the concept of Clausius, a microscopic basis that relates
the entropy of a system to the probability ($p_{i}$) of the macroscopic state based on the
number of equivalent microscopic states which render the macroscopic state indistinguishable,
defining the entropy as: 

\begin{equation}
S=-\sum p_{i}\log p_{i}.
\end{equation}

Entropy, unfortunately, became associated with disorder, the antithesis of complexity and so how was it possible that biological or social systems appeared to increase their complexity and reduce disorder, when the second law demanded its
destruction? One of the most accepted of these explanations establishes that the second law applies 
only to closed systems, and hence it does not apply directly to biological or social systems are 
not closed (they exchange energy, matter and information). Then these kind of systems may only reduce their entropy at the expense of the environment, with which form a closed system. 

As discussed by Brooks in his book Evolution as Entropy, the key to resolve this apparent paradox is to realized that biological and social systems are nonequilibrium phenomena characterized by an increasing phase space and a tendency for realized diversity to lag behind an entropy maximum. In a physical context two of the postulated mechanisms for increasing
the number of micro states were, the expansion of the universe, and the production of elementary particles \cite{Frautschi-1982}.

\begin{figure}
\includegraphics[scale=1]{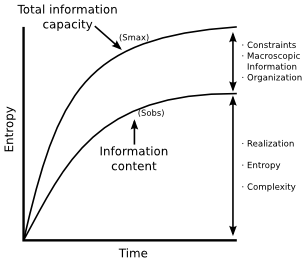}

\caption{Modified from "Evolution as Entropy" \cite{Brooks-1988}, shows the relation between the total information capacity of a system, associated with the maximum available entropy, and the actual information content of the system after all macroscopic constraints are applied.}\label{fig:Diagram}
\end{figure}

Lets consider the diagram in figure \ref{fig:Diagram} where the vertical axis is a measure of the entropy of the system ($S$). If the maximum entropy $S_{max}$ grows with time, then the observed entropy or the final entropy that is attained once all constraints (macroscopical and historical) has been taken into account may also grow. 

Considering a more abstract definition of entropy, given a string of characters
$X$, composed by a sequence of values or symbols $x$ which follow a probability
distribution $P(x)$, information (according to Shannon) is defined
as
\begin{equation}
I=-\sum p(x)\log p(x),
\end{equation}

which has the same functional form of the physical entropy, therefore, in this work we shall talk about entropy ($S$) or information ($I$) as the same. Then the maximum entropy $S_{max}$ is the system total
information capacity while the observed entropy ($S_{obs})$ is the actual information content. The macroscopical constraints are the macroscopical information and the historical constraints are the information historically excluded.

Even more, Gershenson and coworkers \cite{Gershenson-2012} have proposed that complexity ($C$)
may be directly measured as

\begin{equation}
C=aI_{out}(1-I_{out})
\end{equation}

where $a$ is a normalization constant and $I_{out}$ is the system information
after all computations have occurred. In an entropic formulation

\begin{equation}
C=aS_{obs}(1-S_{obs})=aS(1-S).\label{eq:complexity}
\end{equation}

In this form we may measure the complexity of a system by measuring its entropy.

At this point, we present a simple computation in the context of game theory as applied to the social sciences that poses some interesting questions. Essentially we show that if social interactions are described by a game and we compute the associated entropy, then there are some cases in which it will increase monotonically.
More precisely, assume for simplicity and without loss of generality that individuals in a society can adopt two strategies $A$ or $B$ and let $x_{A}$ and $x_{B}$ denote the proportion of the population choosing strategy $A$  and $B$ respectively. So we have $x_{A}+x_{B}=1$. Since they are both positive, we can interpret them as probabilities.
We assume, in the standard context of game theory that they satisfy the replicator equations 

\begin{equation}
\begin{array}{c}
\dot{x}_{A}=\left(f_{A}\left(x_{A},x_{B}\right)-\phi\right)x_{A}\\
\dot{x}_{B}=\left(f_{B}\left(x_{A},x_{B}\right)-\phi\right)x_{B}
\end{array}
\end{equation}

where the dot indicates the derivative with respect to time, and  $f_{A}\left(x_{A},x_{B}\right)$ denotes the fitness level assigned to an individual that adopts the strategy ${A}$ and similarly for the strategy ${B}$, $\phi$ is the average fitness.
Now we may define the entropy of the system as 

\begin{equation}
S(x_{A},x_{B})=-[x_{A}\ln x{A}+x_{B}\ln x_{B}]=-\sum x_{i}\ln x_{i}
\label{eq:entropy}
\end{equation}

and the corresponding entropy production as 

\begin{equation}
\frac{dS(x_{A},x_{B})}{dt}=-\sum[\dot{x}_{i}\ln x_{i}] 
\end{equation}

since $\sum x_{i}=1$ and therefore $\frac{d}{dt}\sum x_{i}=0$. Using the replicator equation we finally get 

\begin{equation}
\frac{dS}{dt}=-\sum x_{i}\left(f_{i}-\phi\right)\ln x_{i}.
\end{equation}

This is an interesting formulation because the sign of the entropy
production depends whether the fitness of a particular population
exceeds the average fitness of the entire population or not. Of course
we should be careful about exactly which entropy we are measuring.
For all open thermodynamic systems the entropy production has two terms, the internal and the external so

\begin{equation}
\frac{dS}{dt}=\frac{dS_{i}}{dt}+\frac{dS_{e}}{dt}.
\end{equation}

All macroscopic systems and processes with an underline physical basis,
are subject to definite thermodynamic laws. On of the primary among these
is the second law of thermodynamics which establishes that the internal
production of entropy due to irreversible processes occurring within
the system, must be positive, 

\begin{equation}
\frac{dS_{i}}{dt}>0.
\end{equation}

and for some special cases where external constraints are constant, classical irreversible thermodynamic theory establishes that the system will eventually arrive at a thermodynamic stationary state in which all macroscopic variables, including the total entropy, are stationary in time, 

\begin{equation}
\frac{dS}{dt}=0,
\end{equation}

and therefore 

\begin{equation}
\frac{dS_{i}}{dt}=-\frac{dS_{e}}{dt}
\end{equation}

implying that

\begin{equation}
\frac{dS_{e}}{dt}<0.
\end{equation}

Maintaining such a systems in a stable thermodynamic stationary state
requires a continuous negative flow of entropy into the system.

As the (Eq\ref{eq:entropy}) is the total entropy of the game, then the game will evolve
to a thermodynamic stationary state when there exists a dominant strategy.
Strategic dominance (commonly called simple dominance) occurs when for one player
one strategy is better than another strategy, no matter
how his opponents may play. Many simple games can be solved
using dominance. The opposite, intransitivity, occurs in games where for one player
one strategy may be better or worse than another strategy, depending on how his opponents may play.

Let us assume that $A$ is dominant strategy and let us write the entropy
explicitly

\begin{equation}
\frac{dS}{dt}=-\left[x_{A}\left(f_{A}-\phi\right)\ln x_{A}\right]-\left[x_{B}\left(f_{B}-\phi\right)\ln x_{B}\right],
\end{equation}

then as $A$ is dominant, the population with this strategy has to decline
over time since its fitness is lower than the average and eventually
will disappear. Then as the population with strategy $B$ is the entire
population, the fitness of $B$ and the average fitness are the same and
the entropy production is zero reaching a stationary state. This of
course means that if the game is representing a social process, this
process will require a continuous negative flow of entropy, which
is not sustainable.

Most interesting is that from the thermodynamic stand point \cite{Michaelian-2000},
 sustainability (understood as the capacity of a system to reach
states of greater longevity) is attainted by minimizing the Helmholtz
free energy

\begin{equation}
F=U-TS\label{eq:helmholtz},
\end{equation}

where $T$ is the temperature of the system associated with the internal
randomness, and $U$ is the internal energy associated with the energy
due to interactions. There are two ways of minimizing Helmholtz free energy, one is by minimizing $U$ and the other is by maximizing $S$. Most of the time the internal energy $U$ is fixed or cannot be controlled externally, so the only alternative is maximizing the entropy.  As $dS/dt>0$ means that the entropy grows and the system is in a more sustainable configuration. In contrast with a $dS/dt<0$, means that the entropy decreases and the system is getting far away from sustainability. 

A positive entropy production is only achievable when the average
fitness is greater that the local one and that corresponds to cooperative games. Cooperation is essential to sustainability, as we all know, but now we have a quantitative indicator for measuring how sustainable is a system. 

Another interesting aspect of this formulation is that, it also makes
clearer the role of entropy in complexity. We have showed that complexity may be indirectly measured by using the system entropy, but again we must be very careful in not oversimplifying this with the false idea that complexity and entropy are exactly the same, they are related but they are not the same. The common conception of complexity, prevalent among physicists, is based on the notion of a noninteracting ideal gas. It is only in this case that a direct association with entropy has some legitimacy. In the real world, interactions are an integral part of the ordering of material particles. In fact, as pointed out by Michaelian (2000) \cite{Michaelian-2000}, a definition of order had been given long ago by Helmholtz in his formulation of the free energy (Eq\ref{eq:helmholtz}). Increasing the entropy in Eq.\ref{eq:helmholtz} decreases the free energy of the system. In this scheme, order has a very natural description, increasing the order of the system makes it more stable, lasting longer in time, more sustainable. This is true of either the ideal gas approaching equilibrium, the crystallization of matter, or the evolutionary dynamics of a community of interacting biological or social systems. 

Then, the association between information and entropy is tricky because complexity is measured with the information after all the computations of the system has been carried out. This information of course has  components from the interactions and then the entropy that must be used is one that includes them, or in other words the entropy that appears in the Helmholtz free energy expression. This allows us to make a direct connection with the entropy production calculated from the replicator equation and identify, as Michaleian \cite{Michaelian-2000} suggests, sustainability with Helmholtz free energy minimization. Then the fitness is a measure of this minimization capacity. 

Until now, in this first approximation, we have only worked with the temporal dimension and
not even fully because we have assumed that the interaction does not
change over time. But, of course, in social systems the interactions
may evolve with time, for example two agents that use to cooperate may
stop suddenly (or vice versa), changing the form of the interaction
and with them the entropy of the system as Axelrod proved \cite{Axelrod-2006}. Even more, the spatial dimension plays a key role in the
emerging of cooperation \cite{Nowak-2006} so, we must understand how it
contributes to entropy.

Lets begin by considering only the spatial contribution to entropy
with non time dependent interactions in a system of $n$ replicator equations in a space defined by a set
of $r_{j}$ coordinates as

\begin{equation}
\frac{\partial x_{i}(r_{j},t)}{\partial t}=x_{i}\left(f_{i}(x_{1},...x_{i},...x_{n})-\phi\right).
\end{equation}

We assume some basic probabilistic properties for this equations as
if $0\leq x_{i}(r_{j},0)\leq1$ then for every subsequent time $0\leq x_{i}(r_{j},t)\leq1$.
In the same way if $\sum x_{i}(r_{j},0)=1$ then for every subsequent
time $\sum x_{i}(r_{j},t)=1$.

Under these considerations the first approximation for considering
a spatial contribution would be by introducing a diffusive term en
the replicator equations

\begin{equation}
\frac{\partial x_{i}(r_{j},t)}{\partial t}=\epsilon_{i}\Delta x_{i}+x_{i}\left(f_{i}-\phi\right),
\end{equation}

where $\epsilon_{i}$ is a diffusion coefficient and $\Delta$ is
the laplacian operator. 

Now a interesting question is, under which conditions $\frac{d}{dt}\underset{\Omega}{\int}S(r_{j},t)dr_{j}$
is bigger than zero, so the second law of thermodynamics holds? 

Again we define the entropy as $S=-\sum x_{i}\ln {x_{i}}$, then

\begin{equation}
\frac{d}{dt}\underset{\Omega}{\int}S(r_{j},t)dr_{j}=\underset{\Omega}{\int}\frac{\partial}{\partial t}S(r_{j},t)dr_{j},
\end{equation}

and using the definition of entropy

\begin{equation}
\frac{d}{dt}\underset{\Omega}{\int}S(r_{j},t)dr_{j}=-\sum\left[\underset{\Omega}{\int}\frac{\partial}{\partial t}\left(x_{i}\ln x_{i}\right)dr_{j}\right].
\end{equation}

Calculating the derivative and using the replication equation

\begin{equation}
\frac{d}{dt}\underset{\Omega}{\int}S(r_{j},t)dr_{j}=-\underset{\Omega}{\int}\left\{ \frac{\partial}{\partial t}\left(\sum x_{i}(r_{j},t)\right)+\sum\left[\epsilon_{i}\Delta x_{i}+x_{i}\left(f_{i}-\phi\right)\right]\ln x_{i}\right\} dr_{j},
\end{equation}

using the normalization condition $\sum x_{i}(r_{j},t)=1$ it turns
out that $\frac{\partial}{\partial t}\left(\sum x_{i}(r_{j},t)\right)=0$,
and then

\begin{equation}
\frac{d}{dt}\underset{\Omega}{\int}S(r_{j},t)dr_{j}=-\sum\underset{\Omega}{\int}\left\{ \left[\epsilon_{i}\Delta x_{i}+x_{i}\left(f_{i}-\phi\right)\right]\ln x_{i}\right\} dr_{j}.
\end{equation}

and finally

\begin{equation}
\frac{d}{dt}\underset{\Omega}{\int}S(r_{j},t)dr_{j}=-\sum\underset{\Omega}{\int}\epsilon_{i}\Delta x_{i}\ln x_{i}dr_{j}-\sum\underset{\Omega}{\int}x_{i}\left(f_{i}-\phi\right)\ln x_{i}dr_{j}.
\end{equation}

Now the sign of the entropy production is, in part, defined by the sign
of $\left(f_{i}-\phi\right)$ in the second integral but, also, by the
spatial term in the first. Analyzing the spatial integral and using
the Divergence Theorem we obtain that

\begin{equation}
\underset{\Omega}{\int}\epsilon_{i}\Delta x_{i}\ln x_{i}dr_{j}=-\underset{\Omega}{\int}\nabla x_{i}\cdot\nabla\left(\ln x_{i}\right)dr_{j}+\underset{\partial\Omega}{\int}\ln x_{i}\frac{\partial x_{i}}{\partial r_{j}}d\sigma;
\end{equation}

where the probability flux is to the inside of the region $\Omega$
when $\frac{\partial x_{i}}{\partial r_{j}}>0$ and outside the
region when $\frac{\partial x_{i}}{\partial r_{j}}<0$. Considering
that $\nabla\ln x_{i}=\frac{1}{x_{i}}\nabla x_{i}$ then we have that
$\underset{\Omega}{\int}\epsilon_{i}\Delta x_{i}\ln x_{i}dr_{j}>0$
for positive probability flux.

These result implies that if $\phi>f_{i}$ and the probability flux
is positive then we recover the second law of thermodynamics $\frac{dS_{i}}{dt}\geq0$. 

Now lets consider a more general where the interaction may change
in time and space. For this instead of the replicator equation, we start by using the L\'opez-Padilla equations \cite{Lopez-2013}.

\begin{equation}
\frac{\partial x_{i}(r_{j},t)}{\partial t}=div\left[e_{i}(r_{j},t)\nabla x_{i}(r_{j},t)\right]
\end{equation}

where $P$ is the probability of finding a player in the position
$r_{j}$ at the time $t$; and $e(r_{j},t)$ is the corresponding
strategy that obeys the equation 

\begin{equation}
\frac{\partial e_{i}(r_{j},t)}{\partial t}=-div\left[f_{i}(r_{j},t)e_{i}(r_{j},t)\right]+\nabla^{2}\left[\phi(r_{j},t)e_{i}(r_{j},t)\right]
\end{equation}

As before we want to calculate the entropy production 

\begin{equation}
\frac{d}{dt}\underset{\Omega}{\int}S(r_{j},t)dr_{j}=-\sum\left[\underset{\Omega}{\int}\frac{\partial}{\partial t}\left(x_{i}\ln x_{i}\right)dr_{j}\right],
\end{equation}

but now we used the new set of equations

\begin{equation}
\frac{d}{dt}\underset{\Omega}{\int}S(r_{j},t)dr_{j}=-\sum\underset{\Omega}{\int}\left\{ div\left[e_{i}(r_{j},t)\nabla x_{i}(r_{j},t)\right]\ln x_{i}\right\} dr_{j}.
\end{equation}

This new expression, for the entropy productions, considers the complete
contributions of space and interactions, but now
an analysis for determine the conditions in which entropy production
is monotonous is not a trivial one, and it will depend on the nature
of the interactions represented by the strategies $e(r_{j},t)$. 

Our main result is that complexity may be measured using entropy and that second law of thermodynamics is recovered under well defined conditions. Our results are very general and takes into account both interaction and space contributions, as well as their time dependence. We have showed that the emergence of cooperation is a consequence of entropic principles and that it may be induce by controlling the difference between mean fitness and the local. A very interesting application of this formalism would be on education.

As we have said, a system is complex when it comprises a large number of interacting subsystems or the problem itself changes with time, whereby it can be concluded that the teaching-learning process is a complex phenomena \cite{Davis-2005, Davis-2006, Frei-2011, Morrison-2006}. On the other hand, research on control of complex systems indicates that managing these dynamic systems can not be achieved through centralized control schemes \cite{Lammer-2008}. On the contrary, a complex adaptive system, such as education, requires to find mechanisms to encourage and guide the processes of self-organization \cite{Argyris-1977}. Thus, a key to improve teaching may be to improve the process per se, but instead design better learning environments (physical and virtual). In this context, the classes may no longer be seen as just teaching exercises but as opportunities to design learning environments that promote connectivity and self-organization, inducing the knowledge to emerge. Our results points that this is achievable by promoting a greater mean fitness value than the local one. These, somehow tell us that an education system based on tests is not coherent with a complex formulation. We should be replacing tests by collaborative activities where the group performance is more important that personal notes. A most interesting experiment, in this directions, has been carried out by Sugata Mitra who, since early as 1982, has been toying with the idea of unsupervised learning and computers. In 1999, he decided to carve a "hole in the wall" that separated the NIIT's premises from the adjoining slum in Kalkaji, New Delhi. Through this hole, a freely accessible computer was put up, for use. This computer proved to be an instant hit among the slum dwellers, especially the children. With no prior experience, the children learned to use the computer on their own \cite{Mitra-2000, Mitra-2001, Mitra-2005}.

Finally, the complexity-based strategies help prepare people for an increasingly complex and intertwined world. Barnett and Hallam (1999) indicate that the world has entered an era of "supercomplexity", characterized by the existence of multiple frames of interpretation \cite{Barnett-1999}. Consequently, knowledge is uncertain, unpredictable, contestable and contested. Against this background, universities must develop autonomy of students, so they can thrive and act deliberately in an uncertain and changing world; therefore, contemporary education should foster skills for: adapting to changes; understanding phenomena in context; making conections between aspects that are not clearly linked; acting against ill-defined situations; dealing with nonlinear behaviors, and in collaboration with others who may not share ideas or interests \cite{Davis-2005, Davis-2006, Frei-2011}.

*{Acknowledgement}
This work was supported by Fondo Capital Semilla at Universidad Iberoamericana; C\'atedras CONACyT program with researcher number 312; SNI numbers: 62929, 14558, 3653; CONACyT Fellowship 471091.

\end{document}